\documentclass[sigconf]{acmart}
\AtBeginDocument{%
  }

\setcopyright{acmlicensed}
\copyrightyear{2026}
\acmYear{2026}
\acmDOI{10.1145/3793302.3793574}
\acmConference[MSR '26]{23rd International Conference on Mining Software Repositories}{April 13--14,
  2026}{Rio de Janeiro, Brazil}
\acmISBN{978-1-4503-XXXX-X/2018/06}

\usepackage{booktabs}
\usepackage{multirow}
\usepackage[most]{tcolorbox}
\usepackage{xurl}
\usepackage{colortbl}

\definecolor{grayish1}{rgb}{0.94902, 0.94902, 0.94902}  %
\definecolor{grayish}{rgb}{0.90196, 0.90196, 0.90196}
\definecolor{yellowish2}{rgb}{1.00000, 1.00000, 0.80000}
\definecolor{yellowish1}{rgb}{1.00000, 1.00000, 0.96078}
\definecolor{yellowish}{rgb}{1.00000, 1.00000, 0.89804}
\definecolor{bluish2}{rgb}{0.80000, 0.80000, 1.00000}
\definecolor{bluish1}{rgb}{0.89804, 0.89804, 1.00000}
\definecolor{bluish0}{rgb}{0.96078, 0.96078, 1.00000}
\definecolor{bluish}{rgb}{0.92549, 0.92549, 0.97255}
\definecolor{greenish1}{rgb}{0.89804, 1.00000, 0.89804}   
\definecolor{greenish}{rgb}{0.89804, 1.00000, 0.89804}
\definecolor{pinkish}{rgb}{1.0000 , 0.8780  , 0.8780 }
\definecolor{ans}{RGB}{243,250,255}

\begin{document}

\title[Insights into  AI-Generated \emph{Silent} Pull Requests]{The Quiet Contributions:\\ Insights into  AI-Generated \emph{Silent} Pull Requests}

\author{S M Mahedy Hasan}
 \affiliation{%
   \institution{Department of Computer Science, Idaho State University}
   \city{Pocatello, ID}
   \country{USA}
   }
 \email{smmahedyhasan@isu.edu}

 \author{Md Fazle Rabbi}
 \affiliation{%
   \institution{Department of Computer Science, Idaho State University}
   \city{Pocatello, ID}
   \country{USA}
   }
 \email{mdfazlerabbi@isu.edu}
  
   \author{Minhaz F. Zibran}
 \affiliation{%
   \institution{Department of Computer Science, Idaho State University}
   \city{Pocatello, ID}
   \country{USA}
   }
 \email{zibran@isu.edu }

\renewcommand{\shortauthors}{Hasan et al.}


\begin{abstract}
We present the \emph{first} empirical study of AI-generated pull requests that are `silent,' meaning no comments or discussions accompany them. This absence of any comments or discussions associated with such \emph{silent} AI pull requests (SPRs) poses a unique challenge in understanding the rationale for their acceptance or rejection.
Hence, we quantitatively study 4,762 SPRs of five AI agents made to popular Python repositories drawn from the AIDev public dataset. We examine SPRs impact on code complexity, other quality issues, and security vulnerabilities, especially to determine whether these insights can hint at the rationale for acceptance or rejection of SPRs.

\end{abstract}

\ccsdesc[500]{Software and its engineering~Software evolution}
\ccsdesc[300]{Software and its engineering~Software quality}
\ccsdesc[300]{Security and privacy~Software security engineering}

\keywords{AI-generated, Pull Requests, Silent, Quiet, Static Analysis, Code Smells, Cyclomatic Complexity, Security Vulnerabilities}



\maketitle

\section{Introduction}\label{sec:intro}
The rapid emergence of AI-assisted software development and code generation has started to reshape the software engineering workflow~\cite{watanabe2025use}. Nowadays, software developers have been increasingly relying on automated coding agents, test bots, and code assistant tools to generate or modify code, often with minimal human involvement, discussions, or reviews. Therefore, this paradigm shift demands a better understanding through critical questions about human-AI collaboration, trust, and software quality issues.

One visible outcome of this shift is the rise of AI-generated \textit{silent pull requests (SPRs)}. We define an AI pull request as \textbf{\textit{SPR}} when it has no comments and is merged or closed without any review comments or discussions. Thus, it is a challenge to determine why such an SPR is accepted or rejected, while the reasons for acceptance and rejection are critical to better understanding how AI agents are performing, which would lead to future improvements.


As there are no comments or discussions associated with SPRs, the only way to explore possible factors behind their acceptance and rejection is to examine the code changes themselves. This motivates our study. We analyze the code before and after each SPR to investigate whether measurable changes in project properties, such as cyclomatic complexity, other code quality issues, or security vulnerabilities, show consistent differences between accepted and rejected cases, which could give us clues about possible reasons.


We, therefore, conduct a quantitative analysis of the impact of 4,762 SPRs on cyclomatic complexity (Research Question 1), other code quality issues (Research Question 2), and security vulnerabilities (Research Question 3), and we examine whether there are differences in the impacts of accepted and rejected SPRs made by five different AI agents. 
We find that SPRs from all agents cause a substantial increase in  complexity and other quality issues, yet they are accepted and merged. Surprisingly, the accepted and rejected SPRs' impacts on the above three criteria, being very similar, remain insufficient to suggest a rationale for their acceptance or rejection. 

This is the \emph{first} empirical study on the \emph{silent AI} pull requests (SPRs).
To facilitate verification and replication of our study, all data and scripts are made publicly available as a replication package~\cite{ReplicationPackage}.

\section{Methodology}
\subsection{Dataset}
\begin{figure}[htbp]
    \centering
    \includegraphics[width=0.45\textwidth,height=0.25\textheight,keepaspectratio]{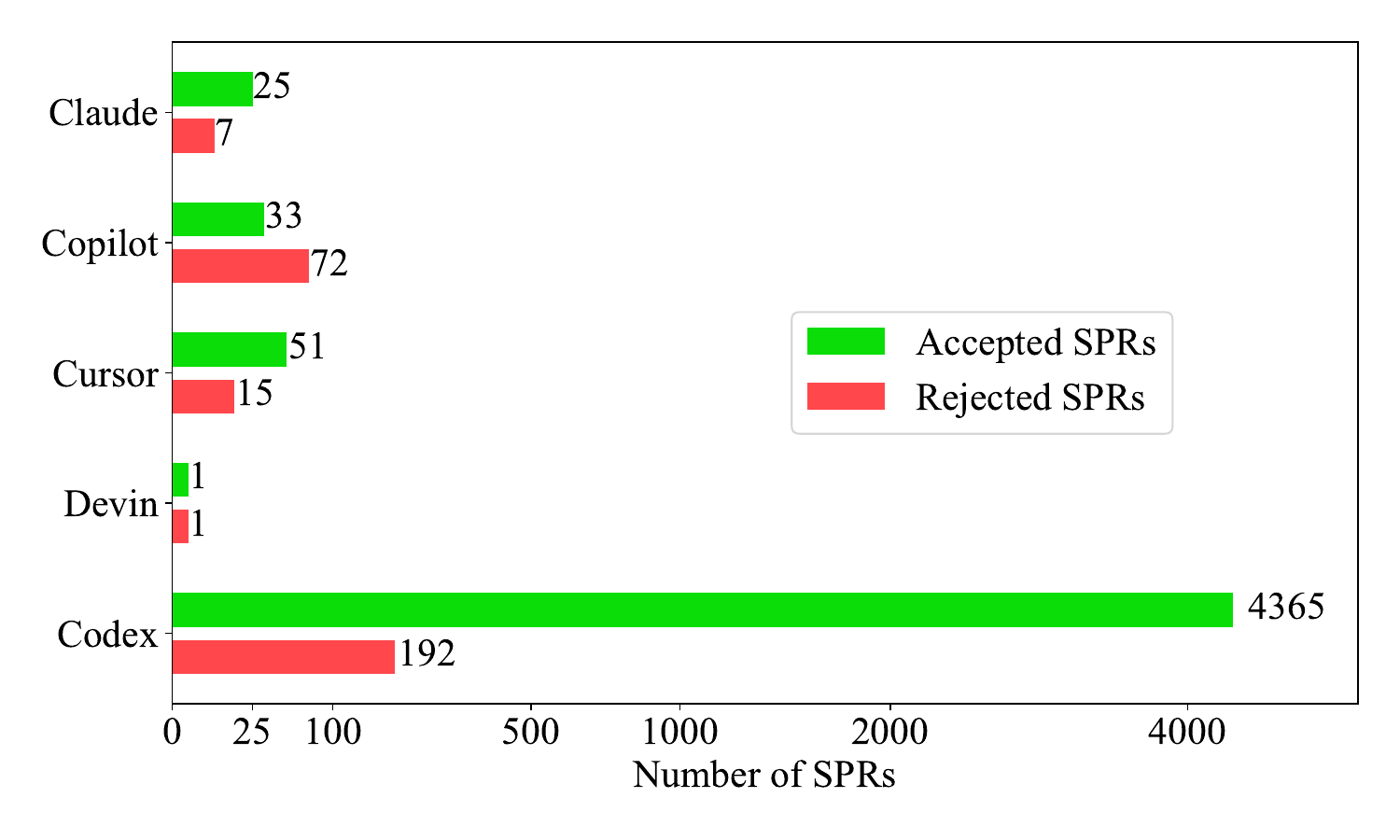}
    \vspace{-0.4cm}
    \caption{Accepted and rejected SPRs from each agent}
    \label{fig:SPR-count}
    \vspace{-0.5cm}
\end{figure}

\vspace{-0.5mm}
For our study, we use the AIDev~\cite{li2025aiteammates_se3} dataset, which is the first large-scale dataset consisting of over 456,000 PRs by five AI agents (i.e., OpenAI Codex, Devin, GitHub Copilot, Cursor, and Claude Code) in 61,000 GitHub repositories.
To derive a manageable subset of this huge dataset, we identify AI PRs for popular Python projects that have more than 100 stars. Thus, we obtain 7,191 Python PRs, of which 5,015 (69.8\%) are SPRs. We focus on Python repositories, because Python is currently among the most popular languages~\cite{Patil2024_ReviewPython} and has the second highest number of pull requests in the AIDev dataset.
Among those 5,015 SPRs, 4,762 are closed (i.e., accepted or rejected) while 253 remain open (i.e., decision pending). All analyses in this study are built on these 4,762 closed SPRs made to the popular Python projects. 
%
%
Figure~\ref{fig:SPR-count} presents the number of accepted and rejected SPRs from each AI agent. 

\subsection{Criteria for Measuring Impact}
To address the research questions outlines in section~\ref{sec:intro}, we measure the impact of SPRs in terms of cyclomatic complexity, other code quality issues, and security vulnerabilities. 

\subsubsection{Computing Cyclomatic Complexity:}
To measure the cyclomatic complexity of a repository, we use Radon 6.0.1~\cite{radon}. Radon statically explores all Python files of the repository and calculates McCabe’s complexity for every function and class, adding them to obtain a project-level complexity value. Radon is widely used in prior studies~\cite{de2025translating, wangkhooklang2025assessing, de2023relationship, hassan2024evaluating, stivala2024investigating, bayram2025comparative, yadav2024revolutionizing, ramirez2025exploring, jamil2025can}. 

\subsubsection{Capturing Code Quality Issues:}
We utilize Pylint 3.0.2~\cite{pylint} to identify code quality issues in four major categories: Error (E), Warning (W), Convention (C), and Refactor (R). 
Pylint, when operated on a repository, examines all Python files and calculates the total number of issues, and also for each issue type, it reports issues by their rule ID and message. 
However, to focus only on important issues that affect code quality, we do not consider minor style-related and docstring warnings captured due to violation of rules C0114, C0115, and C0116. 
Pylint has also been used in many prior work~\cite{rabbi2024ai, thomas2025sustaining, vrechtavckova2025code, jia2024empirical, eghbali2025dylin, xie2024llm, jamil2025can}.

\subsubsection{Identification of Security Vulnerabilities:}
To capture security vulnerabilities, we employ Semgrep 1.68.0~\cite{semgrep}, an open-source static analysis  tool that has been widely used in previous studies to detect security weaknesses in source code~\cite{bennett2024semgrep, kluban2024detecting, johnson2025explaining, munson2025little, aydin2025security} and was reported to have a high (80.49\%) true positive rate~\cite{munson2025little}. It identifies security issues mapped to CWE (Common Weakness Enumeration)~\cite{cwe} identifiers, and each issue is categorized by a CWE-ID and a CWE name. When operated on a repository, Semgrep applies Python security rules to all files and records the total number of CWE-linked issues. 


\subsection{Computational Procedure}
For analyzing the impact of an SPR, we fetch the corresponding repository's snapshot before and after the SPR's final commit, which we call the \emph{before-state} and \emph{after-state} respectively. For accepted SPRs, we use the merge commit to determine the post-integration state of the project (i.e., after-state). For a rejected SPR, as the changes are not integrated with the project, we locally clone the repository and merge the SPR 
to obtain the snapshot of the repository that would be the resultant state of the repository (i.e., after-state) if the SPR were actually accepted and merged. 

For each SPR, we obtain the \emph{before-state} and \emph{after-state} of the associated repository. Then, using Radon, we compute the overall cyclomatic complexity of the before-state and after-state separately. Comparing the overall cyclomatic complexity of the before-state and after-state, we determined whether the complexity increased, decreased, or remained unchanged. Following the same procedure, using Pylint, we determine the impact of an SPR on other code quality issues in terms of whether or not the SPR caused an increase or decrease in code quality issues. Again, we apply the same procedure, using Semgrep, to determine the impact of an SPR on security vulnerabilities.

For code quality and security analysis, we additionally classify issues as \emph{introduced} or \emph{fixed}. A particular issue is considered `introduced' if it appears only in the after-state, and `fixed' if it appears in the before-state only. 


\section{Analysis and Findings}
As seen in Figure~\ref{fig:SPR-count}, the majority (i.e., 4,557 out of all 4,762) of SPRs originated from OpenAI Codex. It is also interesting that 95.7\% of the Codex SPRs are accepted and merged. On the contrary, only one accepted and one rejected SPRs are found that originated from Devin. The effect of these two SPRs from Devin can be considered negligible. Nevertheless, we included them in our analysis for completeness of this study.

\subsection{Impact on Cyclomatic Complexity (RQ1)}\label{sec:complexityAnalysis}
\begin{figure}[htbp]
    \centering
    \vspace{-0.2cm}
    \includegraphics[scale=0.33]{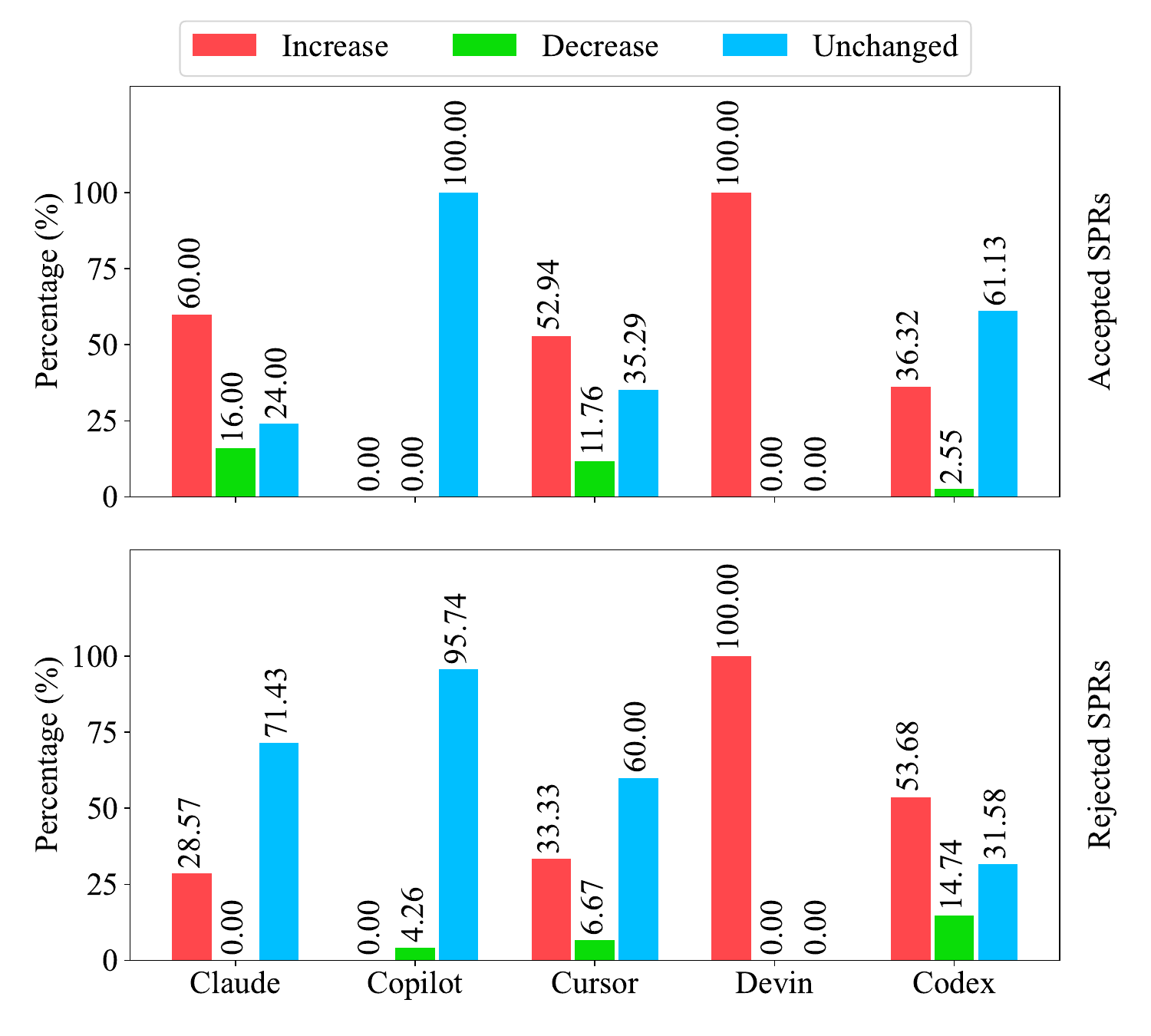}
    \vspace{-0.4cm}
    \caption{AI agents' SPRs' impact on cyclomatic complexity}
    \label{fig:cc analysis}
\end{figure}

To address the research question 1 (RQ1), we first examine the overall effect on cyclomatic complexity of  all SPRs from all AI agents. We find that the majority (59.89\%) of all the SPRs do not affect cyclomatic complexity, while 36.88\% of them cause an increase in complexity, and only 3.23\% of the SPRs reduce complexity. 
To understand whether the impact is uniform across AI agents and especially between SPRs' status (i.e., acceptance or rejection), we further break down the analysis by agent and by SPR status, as displayed in Figure~\ref{fig:cc analysis}. 
The bar chart at the bottom of Figure~\ref{fig:cc analysis} separately presents the percentages of \emph{rejected} SPRs from each AI agent that cause a net increase, decrease, or no-change in cyclomatic complexities. The bar chart at the top presents the same information for the \emph{accepted} SPRs from each AI agent.

As observed in Figure~\ref{fig:cc analysis}, a substantial portion of both accepted and rejected SPRs across all AI agents (except for Devin) leave complexity unchanged.
The SPRs from Devin and Copilot show particularly interesting behaviors. Devin's both accepted and rejected SPRs increase complexity. Inversely, Copilot's all accepted SPRs leave complexity unchanged, while a very small portion (4.26\%) of rejected SPRs slightly decreases complexity. 

In general, all of the AI agents' SPRs (except for Copilot) increase cyclomatic complexity more often than they decrease it, irrespective of their acceptance/rejection status, even though the degree of this effect varies across AI agents. 
If we compare the accepted and rejected SPRs from each AI agent separately, we see that the percentage of Claude's SPRs causing an increase in complexity is substantially higher for accepted SPRs (60.00\%) compared to rejected SPRs (28.57\%). A similar pattern is observed in SPRs from Cursor, but the opposite is observed for SPRs from Codex.
These imply that the SPRs' impact on cyclomatic complexity does not indicate the possibility of their being accepted or rejected.
Based on the above analysis, we now derive the answer to RQ1 as follows:

\vspace{1mm}
\noindent
\colorbox{blue!10}{
\parbox{0.95\columnwidth}{
\textit{\textbf{Answer to RQ1:} An SPR's impact on code complexity does not influence its likelihood of being accepted or rejected. SPRs from all AI agents increase cyclomatic complexity, and complexity-increasing SPRs are more frequent than complexity-decreasing ones. This suggests that integrating AI-assisted changes without discussion can gradually accumulate complexity debt in repositories.}
}}

\subsection{Impact on Other Code Quality Issues (RQ2)}\label{sec:qualityAnalysis}
\begin{figure}[htbp]
    \centering
    \vspace{-0.3cm}
    \includegraphics[scale=0.33]{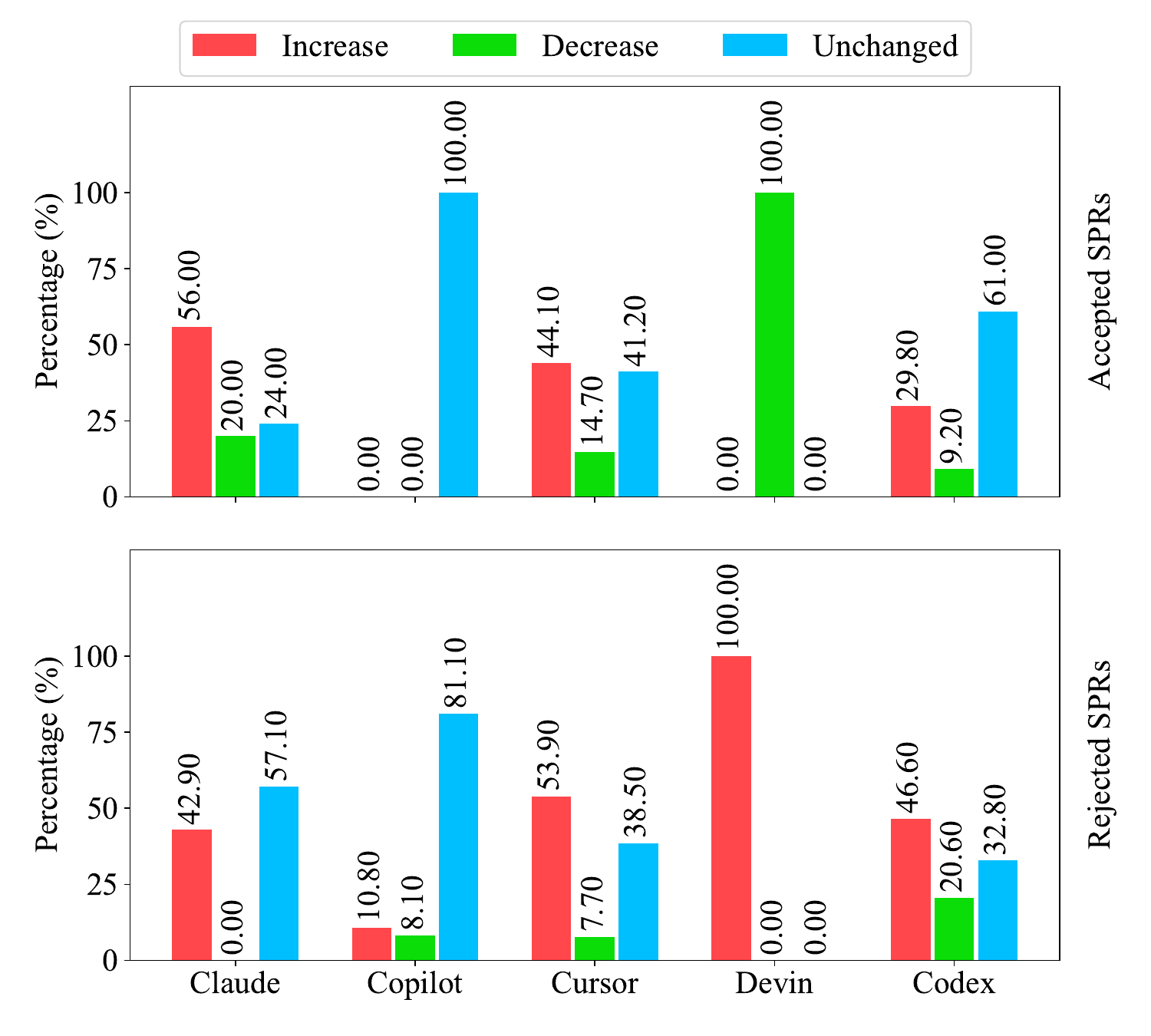}
    \vspace{-0.4cm}
    \caption{AI agents' SPRs' impact on code quality issues}
    \label{fig:pylint code smell}
\end{figure}

To address the research question 2 (RQ2), we first analyze the overall impact of SPRs on code quality analyzing the four categories of issues. Across all agents, we find that the majority (59.70\%) of all SPRs from all AI agents have not changed the number of reported issues in any category. As many as 30.60\% of the SPRs increase the number of code quality issues and only a few (9.70\%) SPRs reduce it.  To understand whether the impact is uniform across SPRs from different AI agents and their accepted/rejected status, we further break down the analysis as portrayed in Figure~\ref{fig:pylint code smell}.

We observe a similar impact of SPRs on other code quality issues as on cyclomatic complexity discussed in Section~\ref{sec:complexityAnalysis}.
As seen in ~\ref{fig:pylint code smell}, with each AI agent (except Devin), and irrespective of acceptance/rejection status, the percentage of SPRs increasing the number of code quality issues is always higher than the percentage of SPRs decreasing code quality issues; the only exception is the accepted SPRs from Copilot, all of which result in no change in code quality issues.
Again, if we compare the accepted and rejected SPRs from each AI agent separately, the percentage of SPRs causing an increase in code quality issues is lower in accepted SPRs (29.80\%) compared to rejected SPRs (46.60\%) from Codex. Similar observation also holds for SPRs from Cursor and Copilot, but surprisingly, the opposite is observed in SPRs from Claude.  

%

Overall, it appears that rejected SPRs contain slightly more issue-increasing modifications, but the pattern is not strong enough to be considered a characterization criterion for distinguishing between accepted and rejected SPRs across AI agents. 


\begin{table}[htbp]
\centering
\small
\caption{Most Frequent Code Quality Issues}
\label{tab:pylint-top-rules}
\vspace{-0.4cm}
\setlength\tabcolsep{3pt}
\small
\begin{tabular}{|l|l|c|r|c|r|c@{}c}
\cline{1-6}
\textbf{AI} & \textbf{SPR} &
\multicolumn{2}{c|}{\textbf{Introduced}} &
\multicolumn{2}{c|}{\textbf{Fixed}} & \multirow{12}{*}{\rotatebox{90}{\footnotesize{Here, rule C0301 refers to ``line-too-long,"}}} & 
\multirow{12}{*}{\rotatebox{90}{\footnotesize{Rule R0801 refers to ``duplicated code."}}} \\
\cline{3-6}
 \textbf{Agent} & \textbf{Status} & \textbf{Rule ID} & \textbf{\%  of SPRs} & \textbf{Rule ID} & \textbf{\% of SPRs} \\
\cline{1-6} \cline{1-6}

\multirow{2}{*}{Claude}
  & Accepted & C0301 & 68.00 & C0301 & 64.00 \\ \cline{2-6}
  & Rejected & R0801 & 71.43 & R0801 & 71.43 \\
\cline{1-6}

\multirow{2}{*}{Copilot}
  & Accepted & R0801 & 25.00 & R0801 & 25.00 \\ \cline{2-6}
  & Rejected & R0801 & 59.46 & R0801 & 62.16 \\
\cline{1-6}

\multirow{2}{*}{Cursor}
  & Accepted & \textbf{C0301} & 44.12 & \textbf{R0801} & 44.12 \\\cline{2-6}
  & Rejected & R0801 & 69.23 & R0801 & 69.23 \\
\cline{1-6}

\multirow{2}{*}{Devin}
  & Accepted & C0301 & 100.00 & C0301 & 100.00 \\\cline{2-6}
  & Rejected & C0301 & 100.00 & C0301 & 100.00 \\
\cline{1-6}

\multirow{2}{*}{Codex}
  & Accepted & R0801 & 93.30 & R0801 & 93.14 \\\cline{2-6}
  & Rejected & R0801 & 57.67 & R0801 & 57.67 \\
\cline{1-6}

\end{tabular}
\vspace{-0.4cm}
\end{table}

\subsubsection{Most Frequent Code Quality Issues:} To better understand the quality issues introduced by SPRs, we identify the most frequently introduced issues. All the 4,762 closed SPRs are found to have introduced 195 distinct issues and fixed 200 different issues. 
An SPR often fixes some quality issues while it also introduces some. Interestingly, despite the large rule space in Pylint for capturing diverse issue types, SPRs, irrespective of their accepted/rejected status, repeatedly fix and introduce a small set of code quality issues. Notably, the most common issues stem from violations of two Pylint rules: C0301 (line-too-long) and R0801 (duplicated code).

In Table~\ref{tab:pylint-top-rules}, we present the prevalence of these two most encountered code quality issues introduced and fixed by the accepted and rejected SPRs from each AI agent. For example, 68.00\% of the \emph{accepted} SPRs from Claude introduce quality issue C0301, while 64.00\% of these SPRs also fixed the same C0301 issue. Surprisingly, the most frequently introduced and fixed issue types and/or percentages often appear to be the same for accepted (or rejected) SPRs from an AI agent. Our manual validation confirms that these are merely coincidences.
%
%
 Based on the above analysis, we now derive the answer to RQ2 as follows:

\vspace{1mm}
\noindent
\colorbox{blue!10}{
\parbox{0.95\columnwidth}{
\textit{\textbf{Answer to RQ2:} Both accepted and rejected SPRs produce more code quality issues than they resolve, and the types of issues they cause are approximately the same across agents. Thus, the impact of SPRs on code quality issues does not hint possibility of their acceptance or rejection.
 }
}}

\subsection{Impact on Security Vulnerabilities (RQ3)}
\begin{figure}[htbp]
    \centering
    \vspace{-0.2cm}
    \includegraphics[scale=0.3]{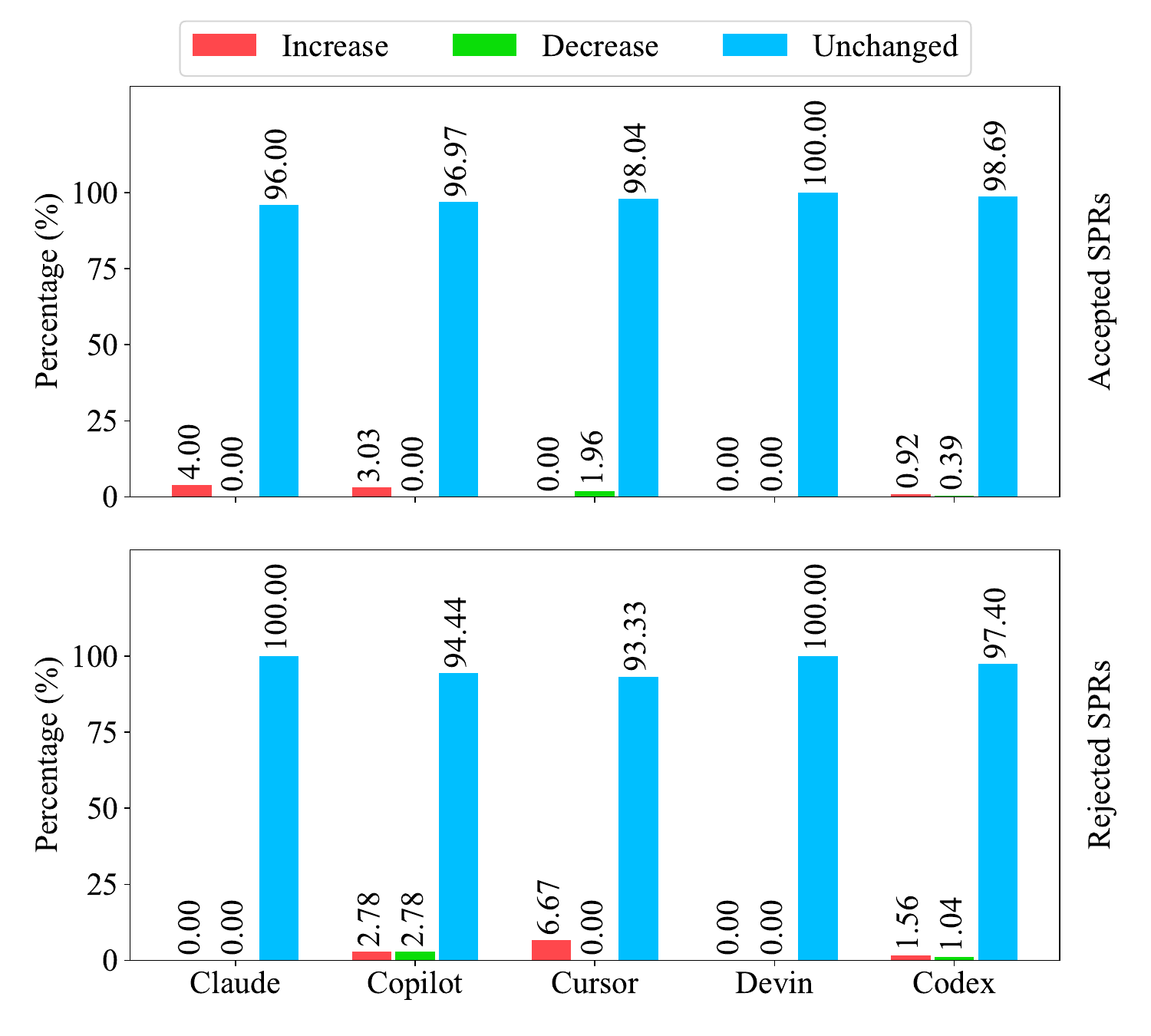}
    \vspace{-0.4cm}
    \caption{AI agents' SPRs' impact on security vulnerabilities}
    \label{fig:semgrep}
    \vspace{-0.1cm}
\end{figure}

 We find that 98.53\% of all SPRs from all agents do not cause a net change in the number of security vulnerabilities in their respective projects. To examine whether the impact is uniform across SPRs from different agents and their acceptance/rejection status, we plot the bar charts in Figure~\ref{fig:semgrep}.

As seen in Figure~\ref{fig:semgrep}, a tiny percentage of \emph{accepted} SPRs from Claude and Copilot cause an increase in the number of vulnerabilities, while 100.00\% of Claude's \emph{rejected} SPRs cause no net change. The percentages of \emph{rejected} SPRs from Copilot and Codex that cause a net increase in vulnerability count are nearly equal to those that cause a net decrease. Only 6.67\% \emph{rejected} SPRs from Cursor cause an increase in net vulnerability count, while even a smaller percentage (1.96\%) of \emph{accepted} SPRs cause a reduction. 
Clearly, most (i.e., 93\%--100\%) of the SPRs from each AI agent, irrespective of their acceptance/rejection state, do not cause any net change in the number of security vulnerabilities in their respective projects.

%


\subsubsection{Most Frequent Security Issues:} 
We find that SPRs introduce 48 distinct CWE IDs and fix 49 unique CWE IDs.  In Table~\ref{tab:semgrep-top-fullpage}, we present the most frequently introduced and fixed CWEs by SPRs from each AI agent as well as the percentage of SPRs that are found to have introduced or fixed them. Detailed description of these CWEs can be found elsewhere~\cite{cwe}.

\begin{table}[htbp]
\centering
\caption{Most Frequent Security Vulnerabilities}
\label{tab:semgrep-top-fullpage}
\vspace{-0.4cm}
\setlength\tabcolsep{3pt}
\small
\begin{tabular}{|l|l|c|r|c|r|}
\hline
\textbf{AI} & \textbf{PR} &
\multicolumn{2}{c|}{\textbf{Introduced}} &
\multicolumn{2}{c|}{\textbf{Fixed}} \\
\cline{3-6}
\textbf{Agent} & \textbf{Status} & \textbf{CWE} & \textbf{\%} & \textbf{CWE} & \textbf{\%} \\
\hline \hline

\multirow{2}{*}{Claude}
  & Accepted & \cellcolor{yellowish2} CWE-532 & 4.00  & \cellcolor{yellowish2} CWE-706 & 4.00  \\\cline{2-6}
  & Rejected & -       & -     & -       & -     \\
\hline

\multirow{2}{*}{Copilot}
  & Accepted & CWE-502 & 9.09  & CWE-502 & 9.09  \\\cline{2-6}
  & Rejected & CWE-489 & 1.39  & CWE-489 & 1.39  \\
\hline

\multirow{2}{*}{Cursor}
  & Accepted & CWE-770 & 9.80  & CWE-770 & 9.80  \\\cline{2-6}
  & Rejected & CWE-502 & 6.67  & CWE-502 & 6.67  \\
\hline

\multirow{2}{*}{Devin}
  & Accepted & CWE-95  & 100.00 & CWE-95  & 100.00 \\\cline{2-6}
  & Rejected & -       & -       & -       & -      \\
\hline

\multirow{2}{*}{Codex}
  & Accepted & CWE-706 & \cellcolor{yellowish2} 1.21  & CWE-706 & \cellcolor{yellowish2} 1.10  \\\cline{2-6}
  & Rejected & CWE-95  & 4.69  & CWE-95  & 4.69  \\
\hline

\end{tabular}
\vspace{-0.3cm}
\end{table}

Surprisingly, SPRs from all agents except Claude, irrespective of their acceptance/rejection status, are found to have frequently introduced and fixed the same vulnerability type (i.e., CWE). Much to our surprise again, we see that, irrespective of acceptance/rejection status, the agent-wise percentage of SPRs that introduce a certain CWE is also equal to that of those fixing the security issue. The only exception is found with accepted SPRs from Codex, where 1.21\% of the SPRs introduced CWE-706, while 1.10\% of those fixed it.
We have manually checked these instances for equal percentages as well as identical security issues, and we conclude that their appearance is also probably coincidental.

It can be observed in Table~\ref{tab:semgrep-top-fullpage} that, across all agents, \emph{accepted} SPRs frequently introduce/fix a relatively diverse set of security issues, including five distinct CWEs (i.e, CWE-95, CWE-502, CWE-532, CWE-706, CWE-770). In contrast, the \emph{rejected} SPRs introduce/fix a smaller set of three CWEs: CWE-95, CWE-502, and CWE-489; however, the first two CWEs (CWE-95 and CWE-502) are also present in the first set.
%
%
%
Based on our observations and above analysis, we now derive the answer to research question 3 (RQ3) as follows:

\vspace{1mm}
\noindent
\colorbox{blue!10}{
\parbox{0.95\columnwidth}{
\textit{\textbf{Answer to RQ3:} Most SPRs from all agents irrespective of their acceptance/rejection status, leave the security posture unaffected and SRPs' impact on security vulnerabilities does not influence the probability of their being accepted or rejected.}
}}

\vspace{-0.1cm}
\section{Related Work}
Prior studies~\cite{tufano2017empirical, tsay2014influence, gousios2014pull, gousios2015work, yu2021reviewer, thongtanunam2017review, rahman2019insight, bavota2015large, silva2016we, palomba2018impact, wattanakriengkrai2020did, rahman2018characterizing, neuhaus2007predicting, han2020finding} on PRs remained focused on \emph{human} PRs and reported that those PRs often introduce code smells, security issues, architectural flaws, and refactoring opportunities, which impact code review time and decisions on acceptance or rejections of the PRs.
While these earlier studies on \emph{human} PRs relied on the associated comments and discussions, we present the \emph{very first} study of those PRs that are \emph{AI-generated} and particularly \emph{silent}, meaning no associated comments and discussions available to extract information from.


\vspace{-0.2cm}
\section{Threats to Validity}
Our work is subject to any limitations of the three widely used tools (Radon, Pylint, and Semgrep). We have studied SPRs' for popular Python repositories only. Thus, our findings may not be generalizable to projects of other programming languages. 


\vspace{-0.1cm}
\section{Conclusion}
In this paper, we have presented the \emph{first} empirical study of \emph{AI-generated silent} pull requests (SPRs).
We have found that, although SPRs cause a substantial increase in cyclomatic complexity and code quality issues, they are still accepted and merged. However, their impact on the number of security vulnerabilities remains limited. Surprisingly, accepted and rejected SPRs exhibit remarkably similar patterns in terms of their impact on the quantity of complexity issues, other quality issues, and vulnerabilities. This similarity suggests that these criteria alone do not explain why some SPRs are merged while others are rejected. 
To find the factors driving these decisions, larger studies are required, which should include additional criteria such as project-specific policies, more diverse software metrics, expertise and historical behavior of code reviewers, as well as contextual cues surrounding non-silent PRs. We plan to extend this work in this direction.
\balance
\bibliographystyle{ACM-Reference-Format}
\bibliography{references}

\end{document}